\documentclass[traditabstract]{aa}

\usepackage{txfonts} \usepackage{graphicx} 
\usepackage{natbib}
\bibpunct{(}{)}{;}{a}{}{,}

\newcommand{\gqcm}{\ensuremath{\mathrm{g} \, \mathrm{cm}^{-3}}}
\newcommand{\kms}{\ensuremath{\mathrm{km} \, \mathrm{s}^{-1}}}
\newcommand{\msol}{\ensuremath{M_\odot}}
\newcommand{\degree}{\ensuremath{^\circ}}
\newcommand{\nhe}{\ensuremath{^{4}\mathrm{He}}}
\newcommand{\ncarb}{\ensuremath{^{12}\mathrm{C}}}
\newcommand{\nnitr}{\ensuremath{^{14}\mathrm{N}}}
\newcommand{\nox}{\ensuremath{^{16}\mathrm{O}}}

\newcommand{\nsul}{\ensuremath{^{32}\mathrm{S}}}
\newcommand{\narg}{\ensuremath{^{36}\mathrm{Ar}}}

\newcommand{\nca}{\ensuremath{^{40}\mathrm{Ca}}}

\newcommand{\nti}{\ensuremath{^{44}\mathrm{Ti}}}
\newcommand{\ncr}{\ensuremath{^{48}\mathrm{Cr}}}
\newcommand{\ncrfn}{\ensuremath{^{52}\mathrm{Cr}}}
\newcommand{\nco}{\ensuremath{^{56}\mathrm{Co}}}
\newcommand{\nfe}{\ensuremath{^{52}\mathrm{Fe}}}

\newcommand{\nni}{\ensuremath{^{56}\mathrm{Ni}}}
\newcommand{\nnin}{\ensuremath{^{57}\mathrm{Ni}}}
\newcommand{\nninn}{\ensuremath{^{58}\mathrm{Ni}}}
\newcommand{\nzn}{\ensuremath{^{60}\mathrm{Zn}}}
\newcommand{\nmo}{\ensuremath{^{98}\mathrm{Mo}}}

\newlength{\percentvaluelen}
\newcommand{\pc}[1]{(\makebox[\percentvaluelen][r]{#1\%})}

\defcitealias{fink2007a}{Paper~I}

\begin{document}

\title{Double-detonation sub-Chandrasekhar supernovae: can minimum
  helium shell masses detonate the core?}

\author{M.~Fink \and
  F.~K.~R\"opke \and W.~Hillebrandt \and I.~R.~Seitenzahl \and
  S.~A.~Sim \and M.~Kromer}

\institute{Max-Planck-Institut f\"ur Astrophysik,
  Karl-Schwarzschild-Str.~1, D-85741 Garching, Germany\\
  \email{mfink@mpa-garching.mpg.de}}

\date{Received 16 December 2009 / Accepted 8 February 2010\rule[-2ex]{0ex}{6ex}}

\abstract{%
  The explosion of sub-Chandrasekhar mass white dwarfs via the double
  detonation scenario is a potential explanation for type~Ia
  supernovae.  In this scenario, a surface detonation in a helium
  layer initiates a detonation in the underlying carbon/oxygen core
  leading to an explosion.
  For a given core mass, a lower bound has been determined on the mass
  of the helium shell required for dynamical burning during a helium
  flash, which is a necessary prerequisite for detonation.  For a
  range of core and corresponding minimum helium shell masses, we
  investigate whether an assumed surface helium detonation is capable
  of triggering a subsequent detonation in the core even for this
  limiting case.
  We carried out hydrodynamic simulations on a co-expanding Eulerian
  grid in two dimensions assuming rotational symmetry.  The
  detonations are propagated using the level-set approach and a
  simplified scheme for nuclear reactions that has been calibrated
  with a large nuclear network.  The same network is used to determine
  detailed nucleosynthetic abundances in a post-processing step.
  Based on approximate detonation initiation criteria in the
  literature, we find that secondary core detonations are triggered for
  all of the simulated models, ranging in core mass from 0.810 up to
  1.385~\msol\ with corresponding shell masses from 0.126 down to
  0.0035~\msol.  This implies that, as soon as a detonation triggers in
  a helium shell covering a carbon/oxygen white dwarf, a subsequent
  core detonation is virtually inevitable.
}


\maketitle

\section{Introduction}
\label{sec:int}

The basic physical mechanism for type~Ia supernova (SN~Ia) explosions
has become widely accepted ever since it was first proposed almost 50
years ago \citep[see e.g.][for a review]{hillebrandt2000a}: a
thermonuclear explosion in electron-degenerate matter
\citep{hoyle1960a} produces radioactive \nni\ that, by its decay,
releases the energy that powers the light curve \citep{truran1967a,
  colgate1969a, kuchner1994a}.  Despite this long history, the
questions of the progenitor system and the explosion scenario have not
been completely answered.  In the so-called double detonation
sub-Chandrasekhar model, a detonation in an accreted helium shell
causes a secondary detonation of a carbon/oxygen white dwarf (C/O WD)
core \citep[e.g.][]{woosley1994b,livne1995a}.  This happens at a total
mass below the Chandrasekhar limit and can lead to a wide range of
possible explosion strengths.  Recent population synthesis studies
suggest that such events are in principle frequent enough to account
for a significant part of the observed SN~Ia rate
\citep{belczynski2005a,ruiter2009a}\footnote{But note that there is
  still some discrepancy between these predictions and those made from
  observations, which tend to be significantly lower
  \citep{deloye2005a,bildsten2007a}.}.

This scenario hinges on two critical points -- first the formation of
a detonation in the helium shell and second whether a successful
detonation of the helium shell can detonate the core.  Here we focus
on the second question.  A secondary detonation can be triggered in
two different ways: either directly when the helium detonation shock
hits the core/shell interface (``edge-lit''), or with some delay,
after the shock has converged near the center.  The edge-lit case is
more restrictive since it requires a strong shock wave and might only
work if the helium detonation starts at some altitude above the base
of the shell \citep[cf.][]{livne1990b}.  In this work we examine the
delayed mechanism since it may still lead to a core detonation even
for shocks too weak for the edge-lit case.  In this spirit of
determining minimal conditions for a core detonation, we look into
models with helium shell masses that are substantially lower than
previously considered.  Provided a detonation is triggered in these,
the shock they drive into the core is expected to be particularly
weak.

\citet[hereafter referred to as Paper I]{fink2007a} found a secondary
core detonation to be robustly triggered in multidimensional
hydrodynamic simulations of generic 1-\msol\ models with shell masses
of 0.1 to 0.2~\msol.  Here we revisit the double-detonation
sub-Chandrasekhar supernova scenario for a series of models with
different core--shell mass combinations.  We use the minimum
helium shell masses required for dynamical runaway calculated by
\citet{bildsten2007a}.  We find that even for this most conservative
case, a helium detonation initiated at the base of the shell robustly
triggers a secondary detonation in the C/O core.  We therefore also
investigate the hydrodynamic evolution, nucleosynthesis, and
observational implications of such explosions.

Section~\ref{sec:expscen} contains the setup of our simulations.  In
Sect.~\ref{sec:numsim} we describe the hydrodynamics and nuclear
reaction network codes used to perform the simulations.  The results
are presented in detail in Sect.~\ref{sec:sim}.
Section~\ref{sec:disc} discusses the results and their observational
consequences in an astrophysical context.  Our work is summarized in
Sect.~\ref{sec:sum}.

\section{Explosion scenario}
\label{sec:expscen}

\subsection{Initial models}
\label{sec:inimod}

The models we study are WDs with a composition of \ncarb\ and \nox\
(equal parts by mass) with minimum accreted helium shell mass
($M_\mathrm{sh}$) that \emph{might} lead to a detonation, as
calculated by \citet{bildsten2007a}.  They assumed a certain
$M_\mathrm{core}$--$M_\mathrm{sh}$ combination and calculated the
resulting structure and $P_\mathrm{b}$, $T_\mathrm{b}$ evolution (with
$M_\mathrm{core}$ being the mass of the accretor WD, and
$P_\mathrm{b}$ and $T_\mathrm{b}$ being the pressure and temperature
at the base of the shell) in a series of hydrostatic equilibrium
integrations assuming a constant-temperature core of $T_\mathrm{core}
= 3 \times 10^7~\mathrm{K}$ and a fully convective shell with an
adiabatic temperature profile.  Heating by the triple-$\alpha$
reaction in a pure helium shell was taken into account.  Thus, the
maximum temperatures during the flash could be determined and also the
shortest nuclear burning timescales.  In series of calculations with
varying $M_\mathrm{sh}$, the minimum flash ignition masses that allow
for a dynamical burning event and potential detonation
$M_\mathrm{min}$ were determined (the term \emph{dynamical} refers to
the condition that the nuclear burning timescale at the shell base,
where it is hottest, becomes as short as the local dynamical
timescale).

\citet{bildsten2007a} also study AM Canum Venaticorum binaries
consisting of a C/O WD accreting at high rates from a helium white
dwarf donor.  They show that for WD masses $\ga$$0.8~\msol$ flash
masses reach or surpass $M_\mathrm{min}$ before the donor mass is
depleted.  This makes these systems interesting candidates for the
sub-Chandrasekhar scenario as they might be frequent and also might
explode at small helium shell masses (0.01--0.1~\msol).

The initial models for our hydro simulations were constructed as
follows: given $\rho_\mathrm{c}$ (central density), $T_\mathrm{core}$,
$\rho_\mathrm{b}$, and $T_\mathrm{b}$ at the instant when burning
becomes dynamical (provided by Bildsten and Shen, private
communication), hydrostatic equilibrium was integrated using our own
code's equation of state.  It differs slightly from
\citet{bildsten2007a}, e.g.\ it does not include Coulomb corrections
to the ion pressure.  Therefore, our core/shell masses might slightly
differ, but this should not change the main properties of the
explosion dynamics.

The most important parameters of our initial models are given in
Table~\ref{tab:inimod}.  In our models we neglect any potential
enhancements of the chemical composition due to the metallicity of the
progenitors.  The core masses $M_\mathrm{core}$ range from
$\sim$0.8~\msol\ for Model~1 to nearly the Chandrasekhar mass for
Model~6.  As already mentioned, the shell masses $M_\mathrm{sh}$ are
set to $M_\mathrm{min}$.  The density at the base of the shell varies
only moderately between 3.7 and $8.7 \times 10^5~\gqcm$.

\begin{table}
  \caption{Parameters of our initial models.}
  \label{tab:inimod}
  \centering
  \begin{tabular}{ccccccc}
    \hline\hline
    Model\rule{0ex}{2.5ex} & 1 & 2 & 3 & 4 & 5 & 6 \\
    \hline
    $M_\mathrm{tot}$ & 0.936 & 1.004 & 1.080 & 1.164 & 1.293 & 1.3885 \\
    $M_\mathrm{core}$ & 0.810 & 0.920 & 1.025 & 1.125 & 1.280 & 1.3850 \\
    $M_\mathrm{sh}$ & 0.126 & 0.084 & 0.055 & 0.039 & 0.013 & 0.0035 \\
    $\rho_{\mathrm{c},\,7}$ & 1.45 & 2.4 & 4.15 & 7.9 & 28 & 140 \\
    $\rho_{\mathrm{b},\,5}$ & 3.7 & 4.0 & 4.5 & 6.1 & 6.4 & 8.7 \\
    \hline
  \end{tabular}
  \tablefoot{$\rho_{\mathrm{c},\,7}$ denotes the central density in
    units of $10^7~\gqcm$, $\rho_{\mathrm{b},\,5}$ is the density at
    the base of the helium shell in units of $10^5~\gqcm$.
    $M_\mathrm{tot}$, $M_\mathrm{core}$, and $M_\mathrm{sh}$ are the
    masses of the WD, the C/O core, and the helium shell,
    respectively.  All masses are given in solar masses.}
\end{table}

\subsection{Detonation initiation}
\label{sec:detcon}

The simulations are carried out in a similar manner as in the previous
study \citepalias{fink2007a}.  In all models, the helium shell is
ignited at a single point at its base.  This one-point ignition
scenario introduces an asymmetry which makes it harder to trigger a
core detonation than in symmetric shell-ignited models. Moreover, it
is expected to introduce viewing angle effects in the synthetic light
curves and spectra.  As the simulation is carried out in 2D rotational
symmetry, the ignition spot is placed on the positive $z$-axis of the
cylindrical coordinate system.

We follow the detonations in the helium shell and in the WD core with
two separate level set functions (see also Sect.~\ref{sec:det}).  The
helium detonation is started by hand by setting the first level set
function to positive values in a volume of choice at the base of the
shell.  Since dynamical burning is a necessary but not sufficient
criterion for the initiation of a detonation, the formation of the
helium detonation is a fundamental assumption in our models.

After shell ignition, the WD core is scanned for ``hot spots'' arising
from the shock exerted on the core by the helium detonation.  If a
sufficiently large volume becomes hot and dense enough, a C/O
detonation is initiated by setting the second level set function to
positive values in this whole volume\footnote{To prevent grid
  geometry effects on the flame, every hot cell initiates a burning
  bubble with a radius of three cells around it.}.  To decide when to
initiate the detonation, critical densities and temperatures given in
the literature are used (\citet{niemeyer1997b,roepke2007a}; see
Tables~1 and 2 of \citetalias{fink2007a}).

These detonation initiation criteria, which are based on the
assumption of spherically symmetric linear temperature gradients
extending into unburned fuel, are limited by some inherent
uncertainties (e.g.\ \citealp{seitenzahl2009b} point out that the
functional form of temperature and density gradients of the hot spot
significantly affects the critical detonation initiation conditions).
Furthermore, the process depends on the composition of the fuel, the
background temperature and local velocity fields\footnote{For such
  effects, volume burning behind the converging shock wave would have
  to be accounted for, but this is not yet implemented in our code.}.
Since the critical length scales pertinent to detonation initiation
are generally much smaller than the grid resolution, addressing these
details is currently impossible in our full-star simulations.
However, the detonation initiations in this work are fairly robust:
the critical temperatures and densities we apply are significantly
exceeded (see Table~\ref{tab:tmax}).  Fully resolved calculations of
the detonation initiation conditions are expected to only moderately
change the critical conditions.  Thus, our conclusions should hold
despite these uncertainties.

\section{Numerical simulation}
\label{sec:numsim}

Our numerical hydrodynamics scheme, briefly summarized below in
Sect.~\ref{sec:hydro}, is similar to the one used in
\citetalias{fink2007a}.  However, there have been major changes to the
treatment of the burning physics, which are described in
Sect.~\ref{sec:det}.  Also, a post-processing scheme is now employed
to determine more realistic nucleosynthetic yields
(Sect.~\ref{sec:pp}).

\subsection{Hydrodynamics}
\label{sec:hydro}

In our Eulerian hydrodynamics code the reactive Euler equations are
solved using a finite volume scheme based on the PROMETHEUS code by
\citet{fryxell1989a}, which is an implementation of the ``piecewise
parabolic method'' of \citet{colella1984a}.  In order to track the
expanding WD during the explosion, a co-expanding uniform grid as in
\citet{roepke2005b}, and \citet{roepke2005c} is used\footnote{The
  expansion of a fixed mass shell is tracked.  To resolve the shock
  convergence arising from the helium detonation, for Models~1--3
  the grid is kept static until the end of helium burning and for
  models~4--6 it is kept static until the onset of C/O detonation.}.
In some of the simulations (Models~4--6) exponentially growing cell
sizes \citep[e.g.][]{reinecke2002d,roepke2006a} are used in the outer
parts of the grid in order to keep a sufficiently high resolution in
the C/O WD during the helium shell detonation and successive shell
expansion.  The grid resolution was $1024 \times 2048$ cells in 2D
rotational symmetry.  An equation of state for WD matter and monopole
gravity complete the system.  The equation of state includes
contributions of an arbitrarily degenerate and arbitrarily
relativistic electron gas, nuclei that follow a Maxwell-Boltzmann
distribution, photons, and electron/positron pairs.  The dependency on
changes in the number of electrons per baryon, $Y_\mathrm{e}$, is also
taken into account.

\subsection{Nuclear burning}
\label{sec:det}

The two detonations are followed by independent level sets in passive
implementation \citep{osher1988a, reinecke1999a, golombek2005a}.  The
burning velocity is calculated as a function of the local density: in
the \nhe\ detonation the Chapman-Jouguet case is assumed and the
burning speed relative to the ashes is the local sound speed.  In C/O
at high densities above $\sim$$2 \times 10^7$ self-sustained
detonations are of pathological type \citep[e.g.][]{sharpe1999a}.  For
the burning speed relative to the fuel the values of
\citet{gamezo1999a} are used above $10^7~\gqcm$.  At lower densities
the Chapman-Jouguet case is assumed.

A simplified scheme is used to model the energy release of nuclear
reactions, similar to \citetalias{fink2007a} and
\citet{reinecke2002b}: the code includes five species variables for
\nhe, \ncarb, \nox, intermediate mass elements (IMEs), and iron group
elements (IGEs).  Changes in composition and internal energy due to
fast reactions occur instantaneously behind the flame discontinuity.
To this end, the final abundances and $Q$-values have been tabulated
against fuel density.  This new prescription constitutes an
improvement to the burning scheme of \citetalias{fink2007a}.  Details
are given in the Appendix.

\subsection{Post-processing}
\label{sec:pp}

Our post-processing scheme is similar to \citet{travaglio2004a}:
passively advected tracer particles are used in the hydro runs and
detailed nucleosynthetic yields are calculated afterwards from their
density and temperature trajectories.  To this end, a large nuclear
network with 384 species ranging from protons, neutrons, and
$\alpha$-particles to \nmo\ is employed.  A description of the code
used to solve the large nuclear network can be found in
\citet{thielemann1990a}, \citet{thielemann1996a}, and
\citet{iwamoto1999a}.  The reaction rate libraries adopted in this
work are the same as in those references, however, newer tables for
weak reaction rates \citep{langanke2000a} were used as described in
\citet{travaglio2004a}.

A number of $n_{\mathrm{core}} = 80 \times 160 = 12\,800$ tracer
particles is placed equidistant in radial mass coordinate and $\cos
\theta$ in the WD core such that each particle represents the same
mass\footnote{An offset is added to the coordinates such that each
  particle has a random initial position within its corresponding
  fluid element.}.  To spatially resolve the nucleosynthesis in the
thin helium shell despite its lower mass, the same number of
$n_{\mathrm{sh}} = n_{\mathrm{core}}$ tracer particles was placed
there.  With this distribution, particles in the core and in the shell
represent masses of $\frac{M_\mathrm{core}}{n_\mathrm{core}}$ and
$\frac{M_\mathrm{sh}}{n_\mathrm{sh}}$, respectively.

\section{Simulation results}
\label{sec:sim}

In this section we present the main simulation results.  All the
models behave in a characteristically similar manner.  Therefore our
discussion will focus on Model~2 ($M_\mathrm{core} = 0.92$,
$M_\mathrm{sh} = 0.084~\msol$) as a detailed example but we will also
comment on the other models where appropriate.

\subsection{Explosive evolution}
\label{sec:gen}

\begin{figure*}
  \sidecaption
  \includegraphics[width=12cm]{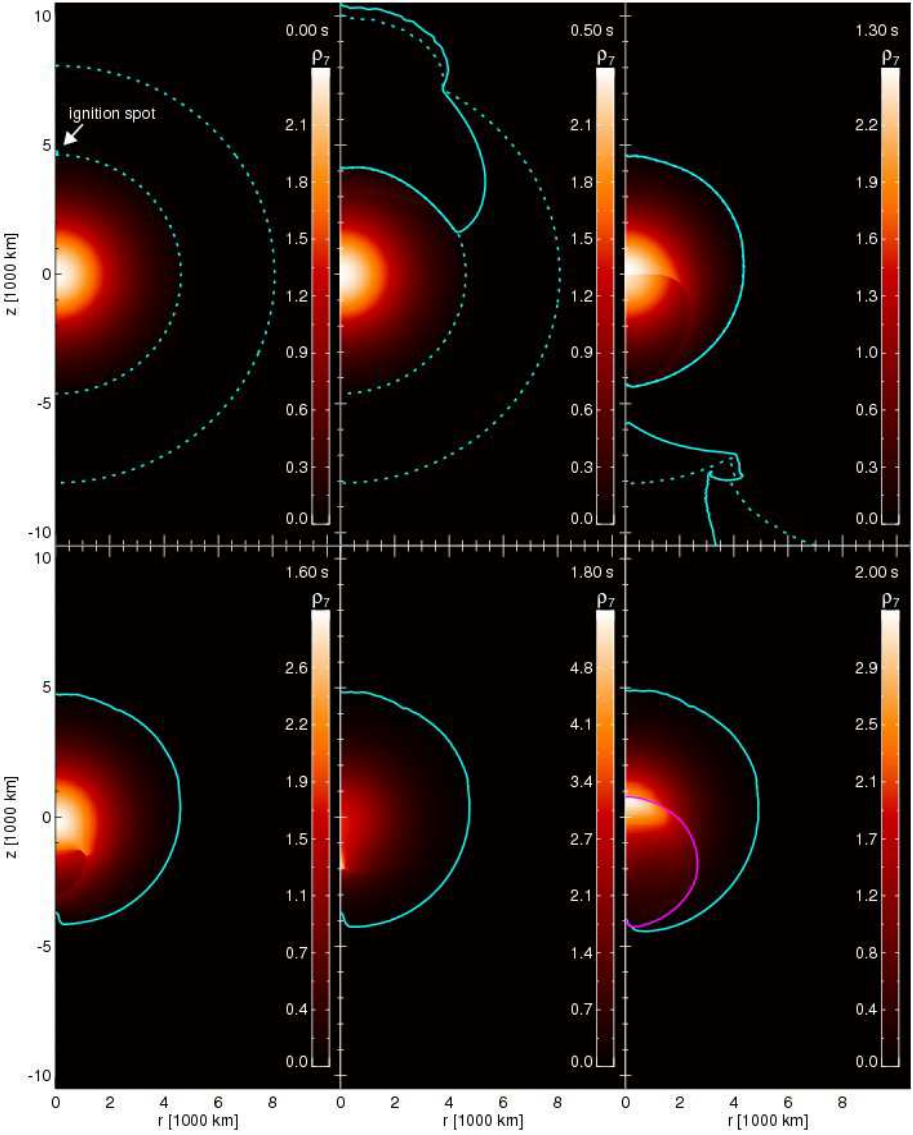}
  \caption{Explosion evolution for Model~2. The density is color
    coded, and the solid cyan and magenta lines are the locations of
    the helium and C/O detonation flames, respectively.  Dashed lines
    in cyan mark the border of the helium shell.}
  \label{fig:mod2}
\end{figure*}

Starting from the ignition point (see top-left panel of
Fig.~\ref{fig:mod2}), the helium detonation surrounds the whole core
until it converges again at the ``south pole''.
While the detonation moves along the base of the shell, an oblique
shock wave propagates into the core, converging off-center at a point
on the negative $z$-axis (cf.\ \citet{livne1990b};
\citetalias{fink2007a})\footnote{Rotation effects that are neglected
  here could break the symmetry of the one-point ignition scenario if
  the ignition spot was located off the rotation axis.  This could
  distort the minimum shocked volume and make a core detonation more
  difficult.  The spherical shell ignition case and all ignition
  geometries that are symmetric with respect to the rotation axis
  should, however, not be influenced.}.  As can be seen in
Fig.~\ref{fig:mod2}, this resembles the self-similar problem of a
spherically or cylindrically converging shock wave
\citep[cf.][]{guderley1942a, landaulifschitz6dt, ponchaut2006a}, which
results in strong shock compression if sufficiently small scales are
resolved.  Due to self-similarity, the scales that need to be resolved
to reach a certain compression just become smaller if the shock from
the shell detonation is weaker.  However, the maximum possible
compression is limited \citepalias[cf.][]{fink2007a}.  The question to
be addressed in this study is whether the volume at which high enough
temperatures and densities for dynamical carbon burning are achieved
is large enough for a detonation to form.  In our simulations we
always reached densities $>$$10^8~\gqcm$ and temperatures $>$$3.2 \times
10^9$~K on resolved scales $>$$2.5$~km.  The critical radius for
detonation formation at these conditions is, however, only 2~m.  Thus,
the critical conditions for detonation initiation are met for all our
models, despite small shell masses.  Table~\ref{tab:tmax} lists the
conditions at which the core detonations were ignited.
\begin{table}
  \caption{Conditions at core detonation initiation for all models.}
  \label{tab:tmax}
  \centering
  \begin{tabular}{ccccccc}
    \hline
    \hline
    Model\rule{0ex}{2.5ex} & $t_\mathrm{ign}$ [s] & $T_\mathrm{9,ign}$ &
    $\rho_\mathrm{8,ign}$ & $z_\mathrm{ign}$ [km] &
    $|z_\mathrm{ign}|/R_\mathrm{core}$ & $\Delta$ [km] \\
    \hline
    1 & 1.94 & 4.62 & 1.08 & $-1910$ & 0.39 & 9.86 \\
    2 & 1.78 & 4.15 & 1.58 & $-1850$ & 0.40 & 8.43 \\
    3 & 1.58 & 4.42 & 2.24 & $-1750$ & 0.41 & 6.81 \\
    4 & 1.30 & 4.09 & 1.89 & $-1560$ & 0.41 & 5.59 \\
    5 & 1.00 & 4.14 & 4.63 & $-1370$ & 0.46 & 3.86 \\
    6 & 0.64 & 3.23 & 8.92 & $-1040$ & 0.50 & 2.54 \\
    \hline
  \end{tabular}
  \tablefoot{$t_\mathrm{ign}$, $T_\mathrm{9,ign}$,
    $\rho_\mathrm{8,ign}$, and $z_\mathrm{ign}$ are the time,
    temperature in $10^9$~K, density in $10^8~\gqcm$, and $z$
    coordinate at the ignition spot, respectively.  $R_\mathrm{core}$
    denotes the radius of the C/O-WD core and $\Delta$ the grid
    resolution, which is approximately $1/1000$ of the WD radius.}
\end{table}
As in \citetalias{fink2007a}, a conservative critical temperature of
$4 \times 10^9$~K was used.  Thus, the given values are only a lower
limit for the maximum possible compression at the given grid
resolution $\Delta$.  If the detonation would have been suppressed,
stronger compression would have been achieved.  Only Model~6 did not
surpass $4 \times 10^9$~K, despite being simulated at the highest
spatial resolution. The conditions reached in the shock compression,
however, were still sufficient to safely assume a successful
detonation triggering.  Therefore, it was ignited at this lower
temperature in a second run.  This model verifies that it is harder to
compress the core sufficiently if the helium shell mass is small.
Conversely, the high initial density of the most massive model makes a
detonation easier.  Based on our approximate initiation criteria we
conclude that the limiting factor for a successful core detonation is
only the successful formation of a detonation in the helium shell.

Table~\ref{tab:tmax} gives times and positions of the detonation
initiations on the $z$-axis.  That the carbon detonations occur
earlier for smaller shell masses can be explained by the decrease of
the core radii associated with the increasing core masses.  At a
smaller radius the helium detonation has a shorter way around the core
while the helium detonation speed at the base of the shell is roughly
constant for all models.  Note that the approximately self-similar
nature of the problem results in the curious fact that the ignition
spots of the core detonations lie at similar relative distances (0.4
-- 0.5 $R_{\mathrm{core}}$) from the center.

\begin{figure}
  \resizebox{\hsize}{!}{\includegraphics{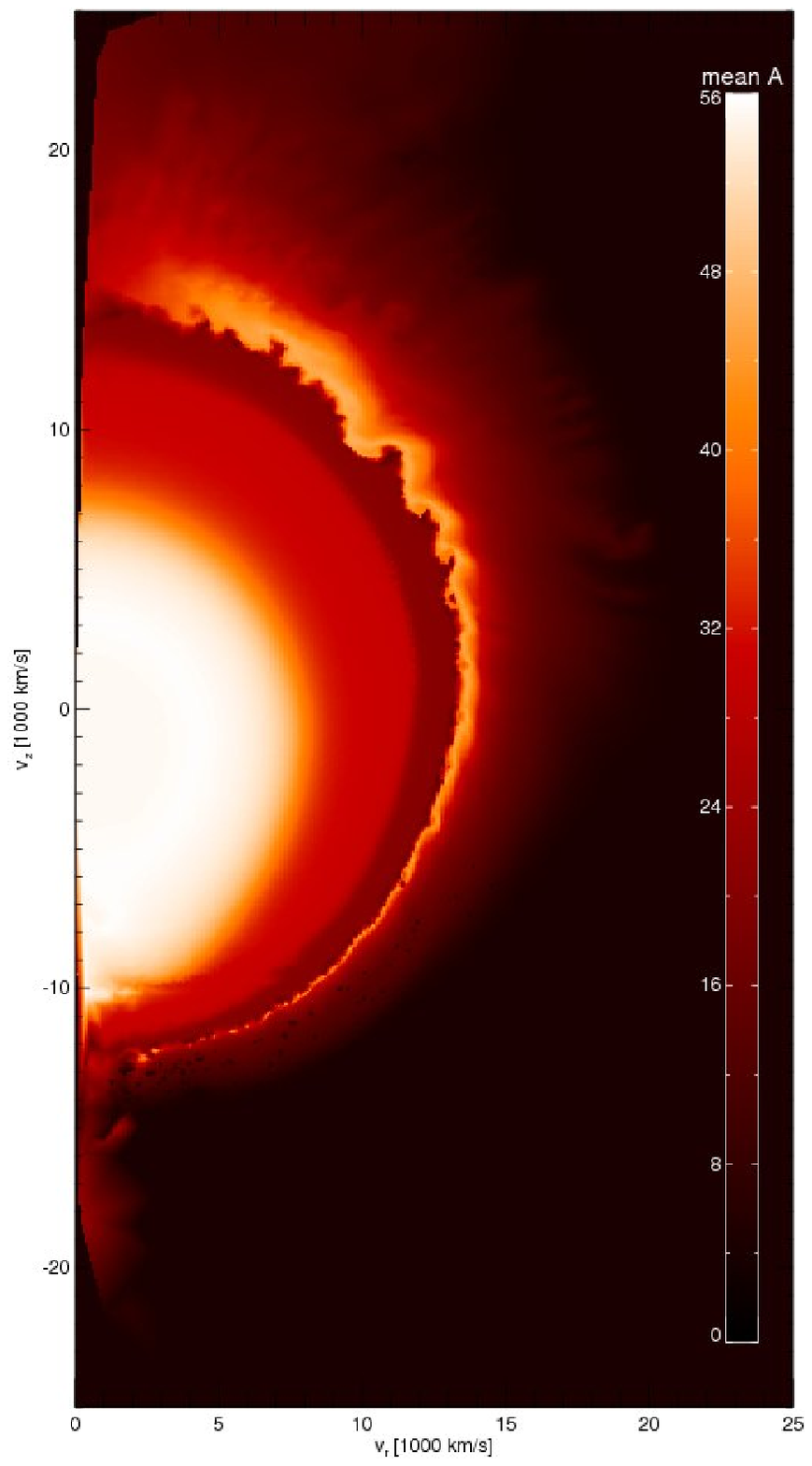}}
  \caption{Ejecta structure in velocity space for Model~2 at 10~s.
    Color coded is the mean mass number (averaged over mass: $\bar{A}
    = \sum_i A_i X_i$, where $A_i$ and $X_i$ are mass number and mass
    fraction of a nucleus $i$, respectively).}
  \label{fig:abar}
\end{figure}

We will now discuss our fiducial case (Model~2) in detail.  At
densities $\la$$4 \times 10^5~\gqcm$ in the helium shell, burning
is relatively incomplete and nuclear statistical equilibrium (NSE) is
not reached.  The final composition is $\sim$63\% of \nhe, $\sim$10\%
of IMEs, and $\sim$26\% of IGEs (see Sect.~\ref{sec:nuc} and
Table~\ref{tab:nuc} for more details).  The C/O detonation starts at
$t \sim 1.8~\mathrm{s}$ and at $z \sim -1900$~km (see
Fig.~\ref{fig:mod2}) and it produces 0.34~\msol\ of \nni, 0.44~\msol\
of IMEs and 0.11~\msol\ of \nox.  The structure of the ejecta at $t =
10~\mathrm{s}$, where our simulations stop and the ejecta are close to
homologous expansion, is shown in Fig.~\ref{fig:abar}.
The shown distribution has been derived from the tracer particles
after the post-processing step.  Due to the two detonations, IGEs can
be found both in the central region and in the shell.  The shock of
the C/O detonation partially penetrates into the helium-detonation
ashes. This gives rise to the Richtmyer-Meshkov instability which
generates vortices at the boundary between C/O- and helium-detonation
ashes (see upper right part of Fig.~\ref{fig:abar}).  This effect
causes some mixing between unburned C/O and IGEs from the helium
detonation.

\subsection{Nucleosynthesis}
\label{sec:nuc}

\begin{figure}
  \resizebox{\hsize}{!}{\includegraphics{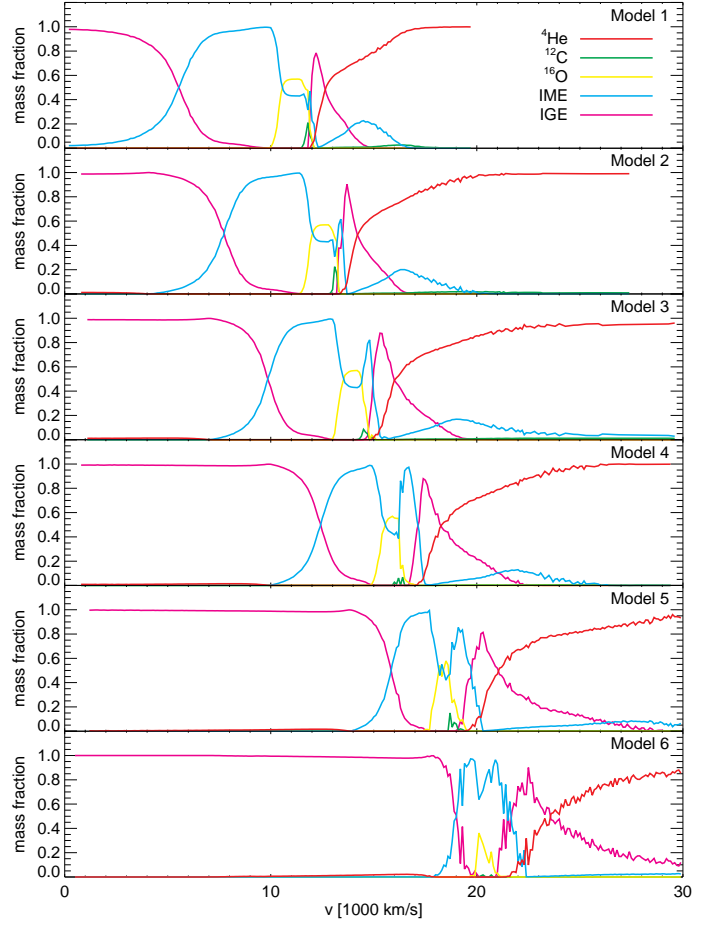}}
  \caption{Mass fractions of the main groups of nuclei in velocity
    space for all models at 10~s.  Shown is the average of an angle
    range of $\pm \pi/32$ around the equator.  The velocity bin size
    is 100~\kms.}
  \label{fig:velprof}
\end{figure}

The total nucleosynthetic yields of the explosions are given in
Table~\ref{tab:nuc} for each model.
\begin{table*}
  \caption{Total nucleosynthetic yields of selected species or groups
    of species.}
  \label{tab:nuc}
  \centering
  \begin{tabular}{lr@{ $\times$ }lr@{ $\times$ }lr@{ $\times$ }lr@{ $\times$ }lr@{ $\times$ }lr@{ $\times$ }l}
    \hline
    \hline

    Model\rule{0ex}{2.5ex} & \multicolumn{2}{c}{1} & \multicolumn{2}{c}{2} & \multicolumn{2}{c}{3} & \multicolumn{2}{c}{4} & \multicolumn{2}{c}{5} & \multicolumn{2}{c}{6} \\

    \hline

          & \multicolumn{12}{c}{C/O core detonation} \\

    $M_\mathrm{C/O,\,fuel}$ & 8.1 & $10^{-1}$ & 9.2 & $10^{-1}$ & \multicolumn{2}{l}{1.03} & \multicolumn{2}{l}{1.13} & \multicolumn{2}{l}{1.28} & \multicolumn{2}{l}{1.39} \\
    $M_\mathrm{IGEs}$       & 1.8 & $10^{-1}$ & 3.6 & $10^{-1}$ & 5.7 & $10^{-1}$ & 8.2 & $10^{-1}$ & \multicolumn{2}{l}{1.11} & \multicolumn{2}{l}{1.33} \\
    $M_\mathrm{IMEs}$       & 4.8 & $10^{-1}$ & 4.4 & $10^{-1}$ & 3.7 & $10^{-1}$ & 2.6 & $10^{-1}$ & 1.2 & $10^{-1}$ & 3.1 & $10^{-2}$ \\
    $M_{\nni}$              & 1.7 & $10^{-1}$ & 3.4 & $10^{-1}$ & 5.5 & $10^{-1}$ & 7.8 & $10^{-1}$ & \multicolumn{2}{l}{1.05} & \multicolumn{2}{l}{1.10} \\
    $M_{\nfe}$              & 7.6 & $10^{-3}$ & 9.9 & $10^{-3}$ & 9.6 & $10^{-3}$ & 7.9 & $10^{-3}$ & 4.2 & $10^{-3}$ & 1.7 & $10^{-3}$ \\
    $M_{\ncr}$              & 3.9 & $10^{-4}$ & 4.6 & $10^{-4}$ & 4.5 & $10^{-4}$ & 3.8 & $10^{-4}$ & 2.1 & $10^{-4}$ & 7.1 & $10^{-5}$ \\
    $M_\mathrm{\nox}$       & 1.4 & $10^{-1}$ & 1.1 & $10^{-1}$ & 8.0 & $10^{-2}$ & 4.2 & $10^{-2}$ & 3.1 & $10^{-2}$ & 1.2 & $10^{-2}$ \\
    $M_\mathrm{\ncarb}$     & 6.6 & $10^{-3}$ & 4.4 & $10^{-3}$ & 2.7 & $10^{-3}$ & 8.8 & $10^{-4}$ & 5.9 & $10^{-3}$ & 7.4 & $10^{-4}$ \\

    \hline

          & \multicolumn{12}{c}{Helium shell detonation} \\

    $M_\mathrm{He,\,fuel}$  & 1.3 & $10^{-1}$         & 8.4 & $10^{-2}$         & 5.5 & $10^{-2}$        & 3.9 & $10^{-2}$         & 1.3 & $10^{-2}$         & 3.5 & $10^{-3}$         \\
    $M_\mathrm{IGEs}$       & 2.9 & $10^{-2}$ \pc{23} & 2.2 & $10^{-2}$ \pc{26} & 1.7 & $10^{-2}$ \pc{30}& 1.3 & $10^{-2}$ \pc{33} & 4.2 & $10^{-3}$ \pc{32} & 1.1 & $10^{-3}$ \pc{31} \\
    $M_\mathrm{IMEs}$       & 1.3 & $10^{-2}$ \pc{10} & 8.2 & $10^{-3}$ \pc{10} & 5.3 & $10^{-3}$ \pc{10}& 5.7 & $10^{-3}$ \pc{15} & 1.9 & $10^{-3}$ \pc{14} & 7.3 & $10^{-4}$ \pc{21} \\
    $M_{\nni}$              & 8.4 & $10^{-4}$ \pc{1}  & 1.1 & $10^{-3}$ \pc{1}  & 1.7 & $10^{-3}$ \pc{3} & 4.4 & $10^{-3}$ \pc{11} & 1.5 & $10^{-3}$ \pc{11} & 5.7 & $10^{-4}$ \pc{16} \\
    $M_{\nfe}$              & 7.6 & $10^{-3}$ \pc{6}  & 7.0 & $10^{-3}$ \pc{8}  & 6.2 & $10^{-3}$ \pc{11}& 3.5 & $10^{-3}$ \pc{9}  & 1.2 & $10^{-3}$ \pc{10} & 2.0 & $10^{-4}$ \pc{6}  \\
    $M_{\ncr}$              & 1.1 & $10^{-2}$ \pc{9}  & 7.8 & $10^{-3}$ \pc{9}  & 4.4 & $10^{-3}$ \pc{8} & 2.2 & $10^{-3}$ \pc{6}  & 6.8 & $10^{-4}$ \pc{5}  & 1.5 & $10^{-4}$ \pc{4}  \\
    $M_{\nti}$              & 7.9 & $10^{-3}$ \pc{6}  & 5.4 & $10^{-3}$ \pc{6}  & 3.4 & $10^{-3}$ \pc{6} & 1.8 & $10^{-3}$ \pc{5}  & 4.9 & $10^{-4}$ \pc{4}  & 6.2 & $10^{-5}$ \pc{2}  \\
    $M_{\nca}$              & 4.7 & $10^{-3}$ \pc{4}  & 3.2 & $10^{-3}$ \pc{4}  & 2.2 & $10^{-3}$ \pc{4} & 2.2 & $10^{-3}$ \pc{6}  & 6.8 & $10^{-4}$ \pc{5}  & 2.4 & $10^{-4}$ \pc{7}  \\
    $M_{\nhe}$              & 8.4 & $10^{-2}$ \pc{66} & 5.3 & $10^{-2}$ \pc{63} & 3.3 & $10^{-2}$ \pc{60}& 2.0 & $10^{-2}$ \pc{52} & 6.9 & $10^{-3}$ \pc{54} & 1.7 & $10^{-3}$ \pc{48} \\

    \hline
                                        
    $M_L$\rule{0ex}{2.5ex} & 2.0 & $10^{-1}$         & 3.7 & $10^{-1}$         & 5.7 & $10^{-1}$        & 8.0 & $10^{-1}$         & \multicolumn{2}{l}{1.06}& \multicolumn{2}{l}{1.10}\\
    $E_\mathrm{kin}$ [$10^{51}$~erg]
                            & \multicolumn{2}{l}{0.90} & \multicolumn{2}{l}{1.04} & \multicolumn{2}{l}{1.20} & \multicolumn{2}{l}{1.40} & \multicolumn{2}{l}{1.59} & \multicolumn{2}{l}{1.68}\\

    \hline
  \end{tabular}
  \tablefoot{$M_\mathrm{C/O,\,fuel}$ and $M_\mathrm{He,\,fuel}$ are
    the total masses of initial fuel in the C/O core and the helium
    shell, respectively.  For the helium detonation the values in
    brackets give the fraction of an isotope mass to the total shell
    mass $M_\mathrm{He,\,fuel}$.  $M_L$ is the total mass of all
    radioactive species that could power a light curve: \nni, \nfe,
    and \ncr.  All masses are given in units of \msol.
    $E_\mathrm{kin}$ is the asymptotic total kinetic energy.}
\end{table*}
The upper part lists the results of the C/O core detonation, whereas
the lower shows those for the helium shell detonation.  Neutronization
becomes important only for the highest mass model showing 16\% of IGEs
that are not \nni.  This model is also peculiar in having almost no
IMEs (only $\sim$2\% of the total mass).  The NSE freeze-out is
$\alpha$-rich for all models showing significant contributions of
\nhe, \nnin, \nninn, and \nzn\ in the final composition.
Fig.~\ref{fig:velprof} shows the distribution of the nucleosynthetic
products in velocity space along the equatorial axis (this is
representative of the mean for the whole explosion).

\begin{figure}
  \resizebox{\hsize}{!}{\includegraphics{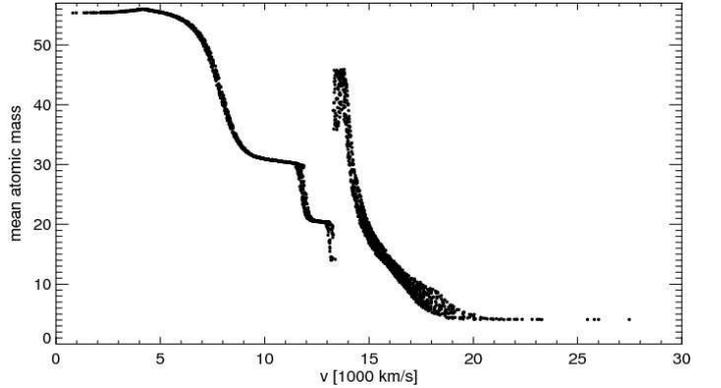}}
  \caption{Mean atomic mass number in the equatorial plane ($z = 0$)
    for Model~2 at 10~s.  Points give the values for individual tracer
    particles within $\theta = \pi \pm \pi/32$.}
  \label{fig:abar_slice}
\end{figure}

A slice of Fig.~\ref{fig:abar} at the $z = 0$ plane gives an overview
of all nucleosynthesis taking place in Model~2
(Fig.~\ref{fig:abar_slice}).
Products of the C/O detonation are located below velocities of
$\sim$13\,000~\kms.  The yields are: IGEs: 39\%, IMEs: 48\%, \nox: 12\%
(see also Table~\ref{tab:nuc}).  For the low central densities present
in most of the models, relatively large amounts of IMEs and \nox\ are
found.  However, there is almost no unburned \ncarb.

\begin{figure}
  \resizebox{\hsize}{!}{\includegraphics{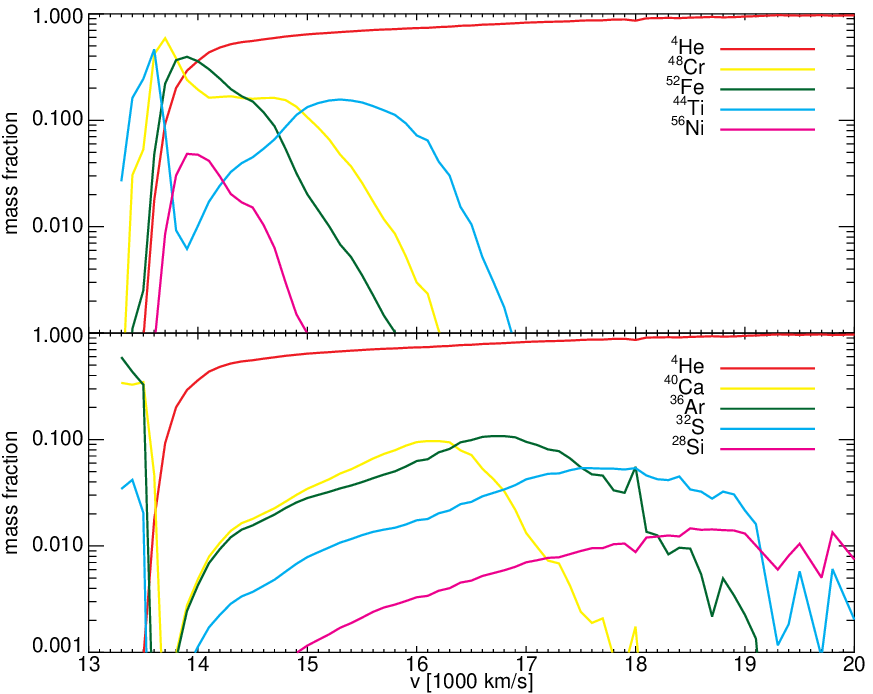}}
  \caption{Distribution of the isotopes in the ejecta in velocity
    space for Model~2 at 10~s.  Shown is the average of an angle range
    of $\pm \pi/32$ around the equator.  The velocity bin size was
    100~\kms.}
  \label{fig:nucleo}
\end{figure}

The helium detonation products are located at $v \ga 13\,000~\kms$
above the C/O detonation layers.  As discussed by
\citet{bildsten2007a}, the burning products at those low densities
differ significantly from previously published values
\citep[e.g.][]{khokhlov1984a, khokhlov1989a}.  For Model~2 the most
abundant nuclei are unburned \nhe\ (63\%), the IGEs \ncr\ (9\%), \nfe\
(8\%), and \nti\ (6\%), and the IMEs \nca\ (4\%), \narg\ (4\%), and
\nsul\ (1\%).  The low-density helium burning regime is characterized
by not reaching NSE and the fact that at the low maximum temperatures
$\alpha$-captures are much faster than the triple-$\alpha$ reactions.
For the low initial densities in the outer shell, helium is mostly
unburned.  Deeper in the shell the higher densities increase the
triple-$\alpha$ rate meaning that more \nhe\ is burned and higher
maximum temperatures are reached.  Once a \ncarb\ nucleus is formed,
$\alpha$-captures process it rapidly to higher mass numbers.  Since
the Coulomb barriers increase with mass number, this process stops at
some maximum mass number depending on the local temperature.  This
leads to an inwards increase of the mean mass number, peaking roughly
at \ncr\ or \nfe.  The final yields of Model~2 in velocity space
(Figs.~\ref{fig:abar_slice} and \ref{fig:nucleo}) clearly demonstrate
this trend.
In the very innermost parts of the shell, where the initial densities
are highest, the mass numbers are again lower.  This is due to
enrichment by carbon produced by triple-$\alpha$ reactions that take
place before the onset of the detonation.  This occurs because of the
high temperatures at the base of the helium shells in our initial
models.  In a mixture of \ncarb\ and \nhe, if the carbon mass fraction
$X_{\ncarb}$ exceeds a value of $\frac{12}{A}$, then there is not
enough \nhe\ to form nuclei with mass number $A$ or higher by
$\alpha$-captures in the course of a detonation passing through this
matter.  E.g., for $X_{\ncarb} \ge \frac{1}{3}$ one expects
$A_{\mathrm{max}} \le 36$.

The initial triple-$\alpha$ burning in the hottest regions of the
shell in the initial models provided by \citet{bildsten2007a} has not
been taken into account in our hydrodynamic simulations, as they lack
a nuclear network that could calculate this volume burning effect.  It
has, however, been considered in our post-processing step.  This
introduces some asymmetry in the results which reflects our choice of
a pure helium shell composition and single spot ignition: while the
detonation wave wraps around the WD, more and more \ncarb\ is produced
in the remaining shell material.  For Model~2, the mass-fraction of
initial \ncarb\ which is reached at the equator by this volume burning
is consistent with the minimum mass fraction values given in
\citet{shen2009a}.  Above the equatorial plane there is less initial
carbon, below there is more.  Therefore our total yields should be
roughly consistent with spherically symmetric or one-dimensional
simulations based on their initial compositions.  The values published
in \citet{bildsten2007a} from the one-dimensional detonation of a
similar model, however, differ from ours as they assume a detonation
in pure \nhe.  Thus it is not surprising that they reach \nni\ by
$\alpha$-captures and find it to be the most abundant burning product.
The fact that they burn $\sim$57\% of the helium and we only about
40\% (in Model~3, which is closest to their model) could be due to the
different explosion dynamics of our one-point ignited and their
whole-shell ignited detonation.

\subsection{Asymmetry effects}
\label{sec:asym}

\begin{figure}
  \resizebox{\hsize}{!}{\includegraphics{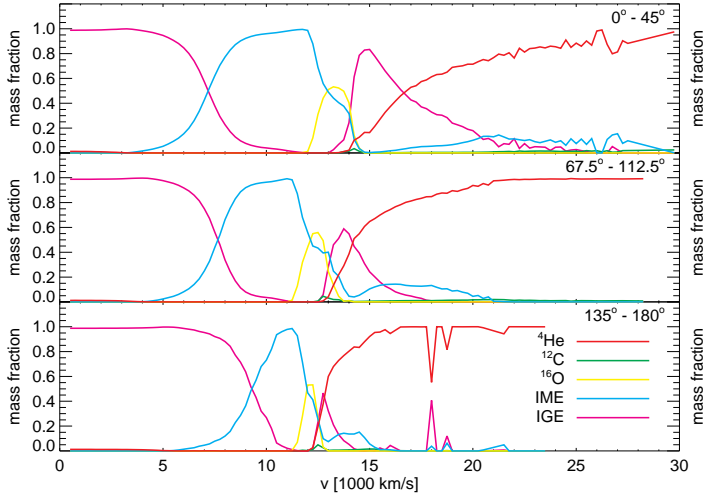}}
  \caption{Asymmetry of mass fractions in velocity space for Model~2
    at 10~s.  (Group) abundances are averaged for three different
    polar angle ranges and over velocity bins of 250~\kms.}
  \label{fig:xasym}
\end{figure}

\begin{figure}
  \centering
  \includegraphics{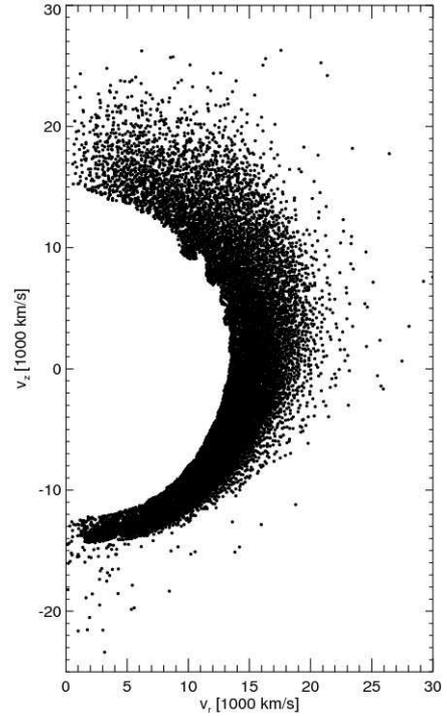}
  \caption{Distribution of tracer particles in velocity space for
    Model~2 at 10~s.  Shown are only particles which initially belong
    to the helium shell.}
  \label{fig:tracers}
\end{figure}

In Fig.~\ref{fig:abar} asymmetries in both composition and ejecta
velocities of Model~2 are visible.  These are quantified in
Fig.~\ref{fig:xasym}, which shows the abundances of the main groups of
elements averaged for three different angular ranges:
0\degree--45\degree\ (``north''), 67.5\degree--112.5\degree\
(``equator''), and 135\degree--180\degree\ (``south'').
The main differences are:
\begin{itemize}
\item Helium detonation products in the north extend over a wider
  velocity range than those in the south.  As the ejecta are already
  homologous to very good accuracy at 10~s, this cannot change by
  stronger expansion of the southern ejecta.  The effect is also
  visible in the final distribution of tracer particles shown in
  Fig.~\ref{fig:tracers}.  As every particle represents the same mass,
  it can also be seen that the southern regions are denser than the
  northern ones.  This property might be important for spectrum
  formation, as a wider range in velocity space increases the range of
  possible photon absorption frequencies.
\item From north to south the amount of burned material decreases.  It
  is 48\%, 37\%, and 24\%, for the northern, equatorial, and southern
  directions of Fig.~\ref{fig:xasym}, respectively.  Also the amount
  of synthesized IGEs is less (corresponding to lower maximum mass
  numbers in Fig.~\ref{fig:abar}): it is 88\%, 66\%, and 51\% of the
  burned matter, respectively.
\item At the south the C/O core has been compressed more strongly: the
  outer boundary of the core ejecta is found to be roughly at
  14\,000~\kms, 13\,000~\kms, and 12\,000~\kms\ from the top panel to the
  bottom panel of Fig.~\ref{fig:xasym}.  This is also consistent with
  the IGE abundance from the core detonation increasing from north to
  south (30\%, 36\%, 55\%) at the expense of IMEs (56\%, 50\%, 38\%)
  and oxygen.
\item While the core ejecta extend to lower velocities in the south
  the IGEs reach significantly higher velocities there.
\item Especially in the northern hemisphere there seems to be
  significant mixing of unburned carbon and oxygen from the core with
  IGEs from the shell detonation.  This might be due to the
  Richtmyer-Meshkov instability (see Sect.~\ref{sec:gen}).
\end{itemize}
Qualitatively similar asymmetry effects appear for the other models of
our study.

\section{Discussion}
\label{sec:disc}

The primary goal of this work has been to investigate whether or not
secondary core detonations can be triggered for scenarios with the
minimum helium shell masses of \cite{bildsten2007a}.  We find that
secondary core detonation conditions leading to a successful explosion
of the WD are obtained in all of our six simulations.  We now comment
on the key observational implications of our double-detonation models.

\begin{figure}
  \resizebox{\hsize}{!}{\includegraphics{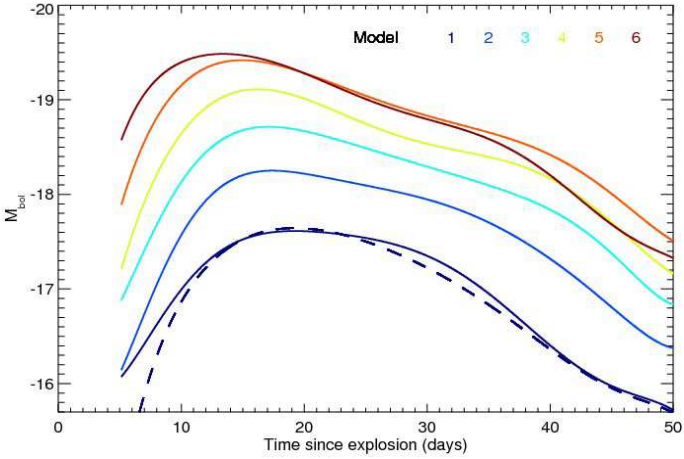}}
  \caption{UVOIR bolometric light curves for the models.  Two light
    curves are shown for Model~1: the first (solid line) shows the
    calculation for the complete model while the second (dashed line)
    shows the result obtained when the contributions of radioactive
    \nfe\ and \ncr\ are neglected.}
  \label{fig:bol_lcs}
\end{figure}

Depending on the initial central density of the model, the C/O
detonation produces nickel masses between 0.17 and 1.1~\msol.  In
principle, this range is sufficiently broad to encompass all major
classes of type~Ia supernovae \citep[cf., e.g.,][for a sample of \nni\
masses determined for 16 well-observed SNe~Ia]{stritzinger2006a}: the
high mass end of sub-luminous
\object{SN~1991bg}\footnote{\citet{filippenko1992a,leibundgut1993a}.}-like
events ($\sim$0.07--0.17~\msol: Model~1), normal type~Ia supernovae
($\sim$0.4--0.8~\msol: Models~2--4), and bright
\object{SN~1991T}\footnote{\citet{phillips1992a}.}-like explosions
($\sim$0.85--1.0~\msol: Models~4, 5).  In Fig.~\ref{fig:bol_lcs} we
show angle-averaged ultraviolet-optical-infrared (UVOIR) light curves
for all six of our models (computed with the {\sc artis} code;
\citealt{kromer2009a,sim2007b}; corresponding values are given in
Table~\ref{tab:bol_lc_paras}).
The light curves illustrate that not only does the peak magnitude vary
significantly between the models, as expected from the differences in
nickel mass, but that there is also significant diversity in both the
rise times (see Table~\ref{tab:bol_lc_paras}) and the post-maximum
light curve shape.
\begin{table}
  \caption{Rise times $t_\mathrm{peak}$ and peak absolute magnitudes
    $M_\mathrm{peak}^\mathrm{bol}$ for the UVOIR bolometric light
    curves of the models.}
  \label{tab:bol_lc_paras}

  \centering
  \begin{tabular}{ccccccc}
    \hline
    \hline

    Model\rule{0ex}{2.5ex} & 1 & 2 & 3 & 4 & 5 & 6 \\
    \hline

    $t_\mathrm{peak}$ [days] & 18.6  & 18.6  & 18.0  & 15.5  & 14.4  & 13.9 \\
    $M_\mathrm{peak}^\mathrm{bol}$   & $-17.6$ & $-18.2$ & $-18.7$ & $-19.1$ & $-19.4$ & $-19.5$ \\
    $\Delta m_{15}^\mathrm{bol}$ & 0.64  & 0.40  & 0.50  & 0.55  & 0.55  & 0.66 \\
    \hline
  \end{tabular}
  \tablefoot{$\Delta m_{15}^\mathrm{bol}$ is the change in bolometric
    magnitude between maximum light and 15~days thereafter.  Due to
    Monte Carlo noise magnitudes are uncertain by $\sim$0.1~mag.}
\end{table}
This diversity arises from the different distributions of the burning
products in both the core and helium shell.  In particular, there is a
clear trend for faster rise times in the brighter models.  This occurs
since, in the brighter models, the \nni-rich core material extends to
higher velocities and the opacity of the outermost layers is less
owing to the lower masses of the IGEs made in the helium shell
detonation.  The fainter models show a single, fairly broad UVOIR
maximum while the brighter models have an initial peak with a weak
shoulder appearing around 30~days after maximum light.  This shape is
qualitatively similar to that obtained for UVOIR light curves for
standard SN~Ia models such as W7 \citep[][]{nomoto1984a}; see figure~7
of \citet{kromer2009a}.  Full details of our radiative transfer
simulations, including complete sets of synthetic light curves and
spectra will be presented in a companion study \citep{kromer2010a}.
  
For most of the models (1--4), the C/O detonation produces a
significant quantity of IMEs ($\sim$0.48--0.26~\msol), as required
to account for the strong lines of e.g.\ silicon, sulphur and calcium
which characterize the maximum-light spectra of SNe~Ia.  The extreme
Model~6, however, makes almost no IMEs and Model~5 yields only a
rather small IME mass ($\sim$0.1~\msol). This means that they are
unlikely to be promising candidates to account for real SNe~Ia.
However, they are still interesting as a demonstration that
detonations of such small shell masses can still trigger a secondary
detonation in a C/O-WD core.  Note that possible initial compositions
favoring more oxygen (and neon) for massive WDs
\citep[cf.][]{dominguez2001a,gil-pons2001a} are not considered here.
Extrapolating from \citet{seitenzahl2009b}, we expect that core
detonations would still be triggered for a composition of 30\% carbon
and 70\% oxygen (in mass).  For compositions with significantly lower
carbon fraction like in oxygen/neon WDs detonation criteria are not
available yet.  Detonation conditions and nucleosynthesis in these
stars should be investigated in future studies.  The concerns
regarding the WD initial composition are alleviated by the fact that
our ignition spots are far above the center (see Sect.~\ref{sec:gen}
and Table~\ref{tab:tmax}) where the carbon fraction is expected to be
higher than in the innermost region \citep[see,
e.g.,][]{althaus2005a}.

The material produced in the helium shell detonation has important
observable consequences.  Although \nni\ is not very abundant in our
models, significant mass-fractions are predicted for the radioactive
nuclei \nfe, \ncr, and \nti.  All of these could play a part in
powering the supernova light curve \citep[cf.\ also][who consider
single helium detonation supernova light curves]{bildsten2007a}.  The
yields of these nuclei are given in Table~\ref{tab:nuc} and, for
Model~1, are as large as 20\% of the \nni\ mass.  The total mass of
all \nni, \nfe, and \ncr\ contributions is given as $M_L$ in
Table~\ref{tab:nuc}.

\nfe\ and \ncr\ have relatively short decay times and release a
similar amount of energy per decay as \nni.  Since they are located in
the outer ejecta (see Fig.~\ref{fig:velprof}), they can contribute to
the early phase light curve of the models in which they are abundant.
This is illustrated in Fig.~\ref{fig:bol_lcs} where, for Model~1, we
compare the light curve computed including the energy released by
\nfe\ and \ncr\ to that obtained if these decay chains are neglected.
Generally, the contribution from \nfe\ and \ncr\ decays is fairly
small and is most significant during the rising phase, as expected.
The light curve at maximum is completely dominated by energy released
from \nni\ and \nco\ decays and it remains so throughout the decay
phase.  There is a modest enhancement around 30~days (of about
$\leq$0.2~mag) that is mostly due to energy produced by $^{48}$V, the
daughter of \ncr\ for which the decay time is $\tau_{1/2} = 16$~days.
The half-life of \nti\ is too long for its decay to directly
contribute to the early light curve.  Nevertheless, the abundance of
titanium is crucial for $U$- and $B$-band light curves and spectra
since even small amounts contribute significantly to the opacity of
the shell.

The high velocity IGEs produced in the helium detonation impose an
important constraint on the ability of our models to reproduce the
early spectra of observed SNe~Ia.  \object{SN~1991bg}-like objects
show titanium in their spectra, but our models are too bright to fit
their light curves.  Normal SNe~Ia do not show clear signatures of
IGEs at early epochs.  Similarly, although there is evidence of IGEs
affecting the pre-maximum spectra of \object{SN~1991T}-like explosions
\citep[e.g.][]{mazzali1995}, these are inconsistent with our models
since the important features there are associated with iron rather
than the IGEs predicted to be abundant from our nucleosynthesis
calculations\footnote{Note that the decay-time of \nfe\ is so short,
  $\sim$0.5~days, that it is expected to have almost completely
  decayed to \ncrfn\ before the light curve is bright.}.  However, to
fully address the issue of whether the presence of heavy elements
created during minimum helium-shell detonations is in contradiction
with observations requires detailed radiative transfer simulations and
consideration of possible uncertainties in the helium-shell
nucleosynthesis.  In particular, we would like to emphasize that the
nucleosynthetic outcome of the helium detonation depends on the
enrichment of the shell with light $\alpha$ nuclei and \nnitr\
\citep[cf.][]{shen2009a}.  Thus, the resulting amounts of \nfe, \ncr,
and \nti\ could be significantly lower than found in this study.  This
will be discussed in a future study \citep{kromer2010a}.

\section{Summary}
\label{sec:sum}

Motivated by the robustness of a secondary core detonation that was
found in a previous study \citepalias{fink2007a}, the
double-detonation sub-Chandrasekhar scenario for SNe~Ia was
re-investigated.  This time we studied the case of minimum shell
masses that might detonate, as calculated by \citet{bildsten2007a} --
this is the most conservative case with regard to the question of
whether secondary core detonations will be triggered.

In order to improve the accuracy of the explosion dynamics over that
of \citetalias{fink2007a}, we switched to a more realistic detonation
prescription in our numerical hydrodynamics simulations, including an
energy release that was calibrated with a large nuclear network.  The
same network was used in a post-processing step to calculate more
detailed nucleosynthetic yields.

Even for shell masses as low as 0.0035~\msol\ and with more realistic
helium detonation physics (not producing pure \nni\ as in
\citetalias{fink2007a}) \emph{a successful core detonation was found
  to be very robust} (for a discussion of uncertainties concerning the
detonation criteria see Sects.~\ref{sec:detcon} and \ref{sec:gen}).
Thus we predict that the ``.Ia''~SNe of \citet{bildsten2007a} turn
into brighter candidates for ``proper'' SNe~Ia.  Given the robustness
of the core detonation and the fact that these events might be
frequent enough to explain a significant part of the SN~Ia rate, the
double-detonation sub-Chandrasekhar scenario is a promising candidate
for explaining some of these thermonuclear explosions.

\begin{acknowledgements}
  We want to thank Lars~Bildsten and Ken~Shen for kindly providing the
  data for our initial models and for helpful discussions.  The
  simulations presented here were carried out at the Computer Center
  of the Max Planck Society, Garching, Germany.  This work was
  supported by the Deutsche Forschungsgemeinschaft via the
  Transregional Collaborative Research Center TRR~33 ``The Dark
  Universe'' and the Emmy Noether Program (RO 3676/1-1) of the German
  Research Foundation.
\end{acknowledgements}

\begin{appendix}

\section{Detonation tables}
\label{sec:app}

\begin{figure}
  \resizebox{\hsize}{!}{\includegraphics{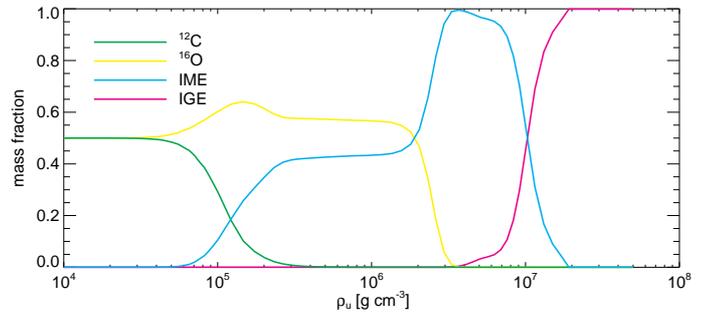}}
  \caption{Mass fractions of the species in the C/O detonation table
    plotted against the density of the unburned fuel
    $\rho_\mathrm{u}$}
  \label{fig:codtab}
\end{figure}

\begin{figure}
  \resizebox{\hsize}{!}{\includegraphics{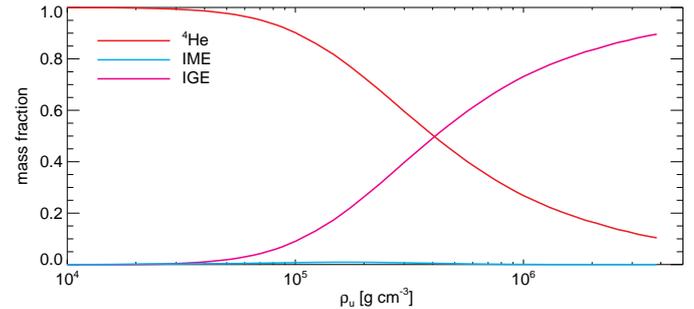}}
  \caption{Mass fractions of the species in the helium detonation
    table against $\rho_\mathrm{u}$.  \ncarb\ and \nox\ abundances are
    not shown, as their values are too close to zero.}
  \label{fig:hedtab}
\end{figure}

In \citetalias{fink2007a} we assumed a certain transition density
between burning to NSE and incomplete burning in the C/O detonation.
Additionally, burning was stopped below a roughly estimated fuel
density.  Therefore the total amount of IGEs and IMEs produced was
relatively uncertain.  For this work we `calibrated' our burning using
a large nuclear network (see Sect.~\ref{sec:pp}) in an iteration of
explosion simulations and post-processing steps: we set up detonations
spherically expanding from the center of a WD star and placed a large
number of tracer particles along one axis.  As starting point, the
production of pure \nni\ at all fuel densities was assumed.  The
density and temperature profiles attained in this way were then used
to determine the detailed nucleosynthetic yields in a post-processing
step (see Sect.~\ref{sec:pp}) with the large network.  Thus the
abundances of our five species (or the respective $Q$-values) could be
tabulated against the initial density of the unburned matter
$\rho_\mathrm{u}$.  This result was then used in another hydrodynamic
simulation of the detonation and a second, more accurate abundance
table was calculated from the post-processing.  After several such
iterations a self-consistent solution was reached.  The resulting
table (see Fig.~\ref{fig:codtab}) is used in all the simulations
presented in Sect.~\ref{sec:sim} and ensures a consistent energy
release and molecular weight in the hydrodynamic simulations.
Analogously a table for detonations in pure helium was determined
(Fig.~\ref{fig:hedtab}).
Although our calibration procedure does not account for potential
changes in the detonation strength in realistic simulations, it still
provides a reasonable approximation.  This was confirmed by
post-processing our simulations presented in Sect.~\ref{sec:sim},
which gave results consistent with the hydrodynamic explosion
simulations.

In the C/O detonation (Fig.~\ref{fig:codtab}) all major burning phases
are visible: carbon burning, oxygen burning, and silicon burning.
Above densities of about $2 \times 10^7~\gqcm$ NSE is reached.  NSE is
represented only by IGEs in this table.  A better representation of
NSE, a temperature and density dependent mixture of \nni\ and \nhe, is
calculated in a different module of the code.  Helium burning
(Fig.~\ref{fig:hedtab}) produces mainly IGEs and only a small amount
of IMEs at the low densities in our shells ($<$$10^6~\gqcm$).  A high
fraction of the initial helium is left unburned.

\end{appendix}

\bibliographystyle{aa} 

\end{document}